%% file: random_20210716_v2_v2_HU.tex
\documentclass[aps, pre, reprint, groupedaddress, showpacs, twocolumn]{revtex4-1}

\usepackage{amsmath,amssymb,bm}
\usepackage{color}
\usepackage{bm}
\usepackage{amsmath}
\usepackage{graphicx}
\usepackage{algorithm}
\usepackage{algorithmicx}
\usepackage{algpseudocode}
\usepackage{afterpage}
\usepackage[normalem]{ulem}
\usepackage[dvipdfmx,
colorlinks=true,
urlcolor=blue,
citecolor=blue,
linkcolor=blue,
hyperfootnotes=false]{hyperref}

\newcommand{\ket}[1]{| #1 \rangle}
\newcommand{\bra}[1]{\langle #1 |}
\newcommand{\braket}[2]{\langle #1 | #2 \rangle}

\begin{document}

\title{Quantum-inspired search method for low-energy states of classical Ising Hamiltonians}

\author{Hiroshi Ueda$^{1,2,3}$}
\author{Yuichi Otsuka$^{2,4}$}
\author{Seiji Yunoki$^{2,4,5,6}$}

\affiliation{$^1$Center for Quantum Information and Quantum Biology, Osaka University,  Toyonaka, 560-0043, Japan}
\affiliation{$^2$Computational Materials Science Research Team, 
RIKEN Center for Computational Science (R-CCS), Kobe 650-0047, Japan}
\affiliation{$^3$JST, PRESTO, Kawaguchi, 332-0012, Japan}
\affiliation{$^4$Quantum Computational Science Research Team, RIKEN Center for Quantum Computing (RQC), Wako, Saitama 351-0198, Japan}
\affiliation{$^5$Computational Condensed Matter Physics Laboratory, RIKEN Cluster for Pioneering Research (CPR), Wako, Saitama 351-0198, Japan}
\affiliation{$^6$Computational Quantum Matter Research Team, 
RIKEN Center for Emergent Matter Science (CEMS), Wako, Saitama 351-0198, Japan}

\date{\today}

\begin{abstract}
We develop a quantum-inspired numerical procedure for searching low-energy states of a classical Hamiltonian composed of 
two-body fully-connected random Ising interactions and a random local longitudinal magnetic field. 
In this method, we introduce infinitesimal quantum interactions that do not commute with the original Ising Hamiltonian, 
and repeatedly generate and truncate direct product states, inspired by the Krylov subspace method, to obtain the low-energy states  
of the original classical Ising Hamiltonian. 
The computational cost is controlled by 
the form of infinitesimal quantum interactions (e.g., one-body or two-body interactions) and 
the numbers of infinitesimal interaction terms introduced, different initial states considered, 
and low-energy states kept during the iteration.
For a demonstrate of the method, here we introduce as the infinitesimal quantum interactions pair products of Pauli $X$ operators 
acting on different sites and 
on-site Pauli $X$ operators into the random Ising Hamiltonian, in which the numerical cost is $O(N^3)$ per iteration 
with the system size $N$. 
We consider 120 instances of the random coupling realizations for 
the random Ising Hamiltonian with $N$ up to 600 and 
search the 120 lowest-energy states for each instance.
We find that the time-to-solution by the quantum-inspired method proposed here, 
with parallelization in terms of the different initial states, 
for searching the ground state of the random Ising Hamiltonian scales approximately as $N^5$ for $N$ up to 600. 
We also examine the basic physical properties such as 
the ensemble-averaged ground-state and first-excited energies 
and the ensemble-averaged number of states in the low-energy region 
of the random Ising Hamiltonian. 
\end{abstract}

\maketitle

\section{Introduction}\label{sec:intro}
There has been renewed interest in quantum annealing 
(QA)~\cite{Kadowaki1998, Kadowaki_arXiv2002, 
Farhi2001, Martonak_PRE2004, Morita2008, Das2008, Ohzeki2011,Albash2018}
particularly because of the successful development of 
the first commercially available real quantum device for QA by D-Wave Systems in 2011~\cite{Johnson_Nature2011}. 
It is indeed noteworthy that the appearance and usage of the QA devices 
have strongly evoked potential needs of solving discrete 
optimization problems for practical applications to  
industrial and social problems.
Although the optimization problem is in general very hard to solve
if we insist on finding the best solution, there are many cases where 
suboptimal solutions can beneficially serve practical purposes, provided
that these solutions are obtained within reasonable computational time.
This is in fact the case for the D-Wave devices because they 
have advantages in finding (near-)optimal solutions much faster than 
classical computers~\cite{Denchev_PRX2016}, 
even though it is difficult to ensure that the best solution is 
always obtained due to finite temperature effects and other sources of noise.

Most of discrete optimization problems can be described as a problem
of finding the ground state of a classical Ising model~\cite{Lucas2014} given by the following Hamiltonian:   
\begin{equation}
\hat{H}_0 = \sum_{i<j} J_{ij} \hat{\sigma}^z_i \hat{\sigma}^z_j + \sum_{i} h_i \hat{\sigma}^z_i,
\label{eq:H0} 
\end{equation}
where $\hat{\sigma}^z_i$ is a Pauli-$Z$ operator for a spin at site $i$.
The problem of finding the ground state of the Ising model for general $J_{ij}$ and $h_{i}$
is NP-hard~\cite{Barahona1982, fu1986, mezard1987}, 
and it is known as the spin glass in statistical physics~\cite{Binder_RMP1986}.

Simulated annealing (SA)~\cite{Kirkpatrick1983} was developed as a numerical 
technique to address this problem, and it exploits thermal fluctuation
to effectively select the ground state among many candidates.
QA is a quantum analogy of SA, 
in which quantum fluctuation instead of thermal one is utilized as driving
force for the annealing process.
It is noted that QA has nowadays also been studied as 
a kind of a quantum adiabatic optimization approach 
in the context of near-term gate-based quantum computers~\cite{Zhou_PRX2020}.
In QA, we introduce an additional Hamiltonian $\hat{H}_1$ 
with $[\hat{H}_0,\hat{H}_1] \neq 0$, 
whose ground state is easily obtained numerically or experimentally. 
We then construct a time-dependent Hamiltonian $\hat{H}(\tau)$ 
within a time interval of $\tau=[0,\tau_0]$,
\begin{equation}
\hat{H}(\tau) = \left(1-\frac{\tau}{\tau_0}\right)\hat{H}_1  + \frac{\tau}{\tau_0} \hat{H}_0,
\label{eq:H_t}
\end{equation}
and consider the real-time evolution
$\ket{\Psi(\tau)} = T_\tau \exp[\frac{1}{i\hbar}\int^{\tau}_{0}\hat{H}(\tau')d\tau']  \ket{\Psi(0)}$,
where $\ket{\Psi(0)}$ is the trivial ground state of $\hat{H}_1$ at $\tau=0$ and
$T_\tau[\cdot]$ denotes the time ordered product.
The advantage of QA is that the adiabatic theorem for quantum 
systems~\cite{Ehrenfest1916,Born1928,Schwinger1937,Kato1950} guarantees
that the ground state of $\hat{H}_0$ is in principle obtained as the final 
state of this process for large enough $\tau_0$, unless there occurs a level 
crossing between the ground state and excited states of $\hat{H}(\tau)$ during 
the entire process.
Even if $\hat{H}(\tau)$ undergoes a first-order phase transition
in the thermodynamic limit, we can modify $\hat{H}(\tau)$ by for example adding 
another quantum Hamiltonian~\cite{Seki2012,Susa2018} to avoid the transition.

While the annealing approaches are usually efficient to search the ground state
of the Hamiltonian $\hat{H}_0$, it should be noted that 
the ground state of the Ising model does not necessarily correspond to the 
best solution of the original optimization problem. 
This is because the mapping from the optimization problem to the Ising model 
is not unique and often involves many hyperparameters. 
The hyperparameters are generally introduced in Eq.~(\ref{eq:H0}) to incorporate 
various types of constraint conditions, 
and are basically tuning parameters that cannot be determined a priori.
Consequently, we need to check whether the ground state estimated by a 
searching algorithm satisfies these constraint conditions; if not, another 
search starting with different values of the hyperparameters is required. 
Considering that there is no general way to tune the hyperparameters, 
this repeating process is sometimes computationally demanding.

If we can obtain a number of low-energy states of the Ising model at once
instead of repeating the single search, the above process can be considerably simplified, 
because one might expect that the best solution satisfying the constraint conditions is still 
found in the low-energy excited states of the mapped Hamiltonian $\hat{H}_0$.
A Monte Carlo method such as the histogram reweighting 
techniques~\cite{Ferrenberg1988,Ferrenberg1988err} and 
the Wang-Landau sampling~\cite{Wang2001} 
might be a good candidate for this purpose.
Indeed, these methods can efficiently evaluate the density of states of a classical 
Ising model in a wide range of energy. 
However, they are not suitable for accurately determining the sequence of 
lowest-energy states, for example, the lowest 100 excited states from the ground state. 
Therefore, it is highly desirable to develop a numerical solver for 
finding a number of low-energy states of $\hat{H}_0$ 
with computational complexity of $O\left({\rm poly}(N)\right)$.

In this paper, we propose a quantum-inspired (QI) method for finding a number of 
low-energy states of the classical Ising  Hamiltonian $\hat{H}_0$ in 
Eq.~(\ref{eq:H0}) with random couplings $J_{ij}$ and $h_i$. 
In this algorithm, 
we generate classical direct product states
by introducing an infinitesimal quantum interaction $\hat{H}_1$
and subsequently truncate those sates with higher energies, which constitutes a single iteration. 
We then repeat this procedure until the predetermined condition is satisfied to obtain the desired number of lowest-energy states. 
This algorithm is inspired by the Krylov subspace method~\cite{KSM}, known as 
one of the most successful algorithms in the numerical linear algebra. 
The computational cost and accuracy are controlled by 
the choice of $\hat{H}_1$ and the numbers
of iterations, 
initial classical states, and 
states kept. 
For a demonstration, we consider all possible pair products of Pauli $X$ operators acting on different sites 
and on-site Pauli $X$ operators as $\hat{H}_1$,  
and show that the computational cost scales as $N^3$ per iteration with $N$ being 
the total number of Ising spins. 
We take 120 different instances of the random coupling $J_{ij}$ and $h_i$ realizations 
for each $N$ up to 600, and search the 120 lowest-energy states for each random coupling instance, 
to examine the scalability of the proposed QI search method. 
Additionally, we discuss the basic properties of the random Ising model such as 
the ensemble-averaged ground-state and first-excited energies and the ensemble-averaged number of states in 
the low-energy region of $\hat{H}_0$.

The rest of this paper is organized as follows. 
We first introduce the physical background of the QI search method proposed here in Sec.~\ref{sec:phys_bg_qi_search}  
and describe the algorithm for the search of low-energy states of the classical Ising model 
$\hat{H}_0$ in Sec.~\ref{sec:algorithm}. 
We then analyze the efficiency of the algorithm 
for searching the 120 lowest-energy states, including the ground state, of $\hat{H}_0$ in Sec.~\ref{sec:results}.
We also discuss the basic physical properties of $\hat{H}_0$ 
obtained by the QI search method and other methods in Sec.~\ref{Sec:H0}. 
The summary and conclusion are finally provided in Sec.~\ref{sec:summary}. 
The performance of the QI search method and the parallel-tempering Monte Carlo (PTMC) method is compared in 
Appendix~\ref{sec:estimation_quality}.   

\section{Physical background of the Quantum-inspired (QI) search method}
\label{sec:phys_bg_qi_search}

The QI search method can be formulated as a classical limit of the $M$-block two-step 
Krylov subspace method, which is a variant of the two-step Lanczos method~\cite{Hieida1997}. 
In this section, first we briefly review the $M$-block two-step Krylov subspace method 
to calculate the $M$ lowest eigenenergies and the corresponding eigenstates of a given 
Hamiltonian, 
\begin{equation}
\hat{H}_{\rm tot} = \hat{H}_{0} + \lambda \hat{H}_{1},
\end{equation}
where $\hat{H}_{0}$ and $\hat{H}_{1}$ are noncommuting classical and quantum 
Hamiltonians, respectively, as is introduced in Sec.~\ref{sec:intro}, and $\lambda$ is a coupling constant.
We assume the eigenvalues $E_\psi$ of $H_0$ in the ascending order with $0 \leq \psi < 2^N$ 
and the corresponding classical eigenstates $\ket{\psi}$, i.e., 
$\hat{H}_{0} \ket{\psi} = E_\psi \ket{\psi}$.

In the $M$-block two-step Krylov subspace method, 
we first prepare $M$ random orthonormal classical states 
$\{ \ket{i}:=\ket{\psi^{~}_i} \}_{1\leq i \leq M}$ 
with $0 \leq \psi^{~}_i < 2^N$, 
which are eigenstates of $\hat{H}_0$, i.e., 
\begin{equation}
\hat{H}_0 \ket{i} = E_{\psi_{i}} \ket{i},
\end{equation}
with the eigenvalues $E_{\psi_{i}}$ in the ascending order:  
$E_0 \leq E^{~}_{\psi^{~}_{1}} \leq \cdots \leq E^{~}_{\psi^{~}_{M}}$. 
Next, we generate $M$ quantum states,
\begin{equation}
\ket{\phi^{~}_i} := \hat{H}_{\rm tot} \ket{i}, 
\label{eq:two-step}
\end{equation}
and 
using, e.g., the modified Gram-Schmidt orthonormalization, 
we orthonormalize these resulting states $\ket{\phi^{~}_i}$ as well as the states $\ket{i}$ 
to obtain the following $2M$ orthonormal states:
\begin{equation}
\ket{\phi^{\prime}_j} := \left\{
\begin{matrix}
\ket{j} & 1 \leq j \leq M \\
\ket{\tilde{\phi}^{~}_{j-M}} & M+1 \leq j \leq 2M \\
\end{matrix} \right.,
\end{equation}
where the state $\ket{\tilde{\phi}^{~}_{j-M}}$ is proportional to $\ket{\phi^{~}_{j-M}} - \sum_{j'<j} \braket{\phi^{\prime}_{j'_{~}}}{\phi^{~}_{j-M}} \ket{\phi^{\prime}_{j'_{~}}}$.

After the orthonormalization, we next construct 
the following $2M$-dimensional effective Hamiltonian:  
\begin{equation}
\hat{H}^{\prime}_{\rm tot} := \sum_{j,j'=1}^{2M} \bra{\phi^{\prime}_j} \hat{H}_{\rm tot} \ket{\phi^{\prime}_{j'}} \ket{\phi^{\prime}_j}\bra{\phi^{\prime}_{j'}},
\end{equation}
and diagonalize this Hamiltonian $\hat{H}^{\prime}_{\rm tot} $ 
to obtain the $2M$ eigenenergies $\{E^{\prime}_j\}$ in the acending order with the 
corresponding eigenstates $\{ \ket{\Phi^{~}_j} \}$, i.e., 
\begin{equation}
\hat{H}^{\prime}_{\rm tot} \ket{\Phi^{~}_j} = E^{\prime}_j \ket{\Phi^{~}_j}.
\label{eq:tsEigen}
\end{equation}

We then set $\ket{i} := \ket{\Phi^{~}_i}$ for $i=1,2,\dots,M$ as the initial states in Eq.~(\ref{eq:two-step})
and repeat the above procedure in Eqs.~(\ref{eq:two-step})-(\ref{eq:tsEigen}) 
until the $M$ lowest eigenenergies $E_j^\prime$ converge within the desired accuracy. 
In this way, the $M$-block two-step Krylov subspace method is able to compute 
not only the ground-state energy but also the low-lying excited-state energies, 
which is the advantage over the standard (one-block) Krylov subspace method. 
The $M$-block two-step Krylov subspace method can be applied to any quantum system
if one can store $2M$ vectors of length $2^N$.
However, it is obviously impossible for large $N$ 
since the dimension of vector grows exponentially with $N$,
resulting in the memory bottleneck.   
Therefore, we need an approximation for large $N$ and 
here we consider a classical limit of the $M$-block two-step Krylov subspace method 
using only classical basis sets, 
where merely $O(M \sim {\rm poly}(N))$ coefficients are required to keep in the memory.

As in the $M$-block two-step Krylov subspace method, 
we first prepare the orthonormal $M$ initial classical states  
$\{ \ket{i} \}$.
We then expand the set of classical states 
by following the guiding principle of the Krylov subspace method, 
i.e., 
\begin{equation}
\{ \ket{\psi'_k} \}^{~}_{1 \leq k \leq K} := \{ \ket{\psi}~;~  \bra{\psi} \hat{H}^{~}_{\rm tot} \ket{i}  \ne 0\}.
\label{eq:hprod}
\end{equation}
In a typical system in condensed-matter physics,  
including the Ising model described by the Hamiltonian in Eq.~(\ref{eq:H0}), 
the number of $K$ is up to $O(N^2)$ because the quantum Hamiltonian $\hat{H}_{1}$ 
consists of only local $n$-body spin flip terms with $n$ typically up to 2. 
Here we assume that the classical states $\ket{\psi'_k}$ with $\psi'_k \in [0,2^N)$ are 
ordered ascendingly with respect to $k$, i.e., $E^{~}_{\psi^{'}_{1}} \leq E^{~}_{\psi^{'}_{2}} \leq \cdots \leq E^{~}_{\psi^{'}_{K}}$.
Next, these $K$-dimensional basis states are used to construct 
the following effective Hamiltonian:  
\begin{equation}
\hat{H}^{\prime}_{\rm tot} := \sum_{k, k'=1}^K \bra{\psi^{\prime}_k} \hat{H}_{\rm tot} \ket{\psi^{\prime}_{k'}} \ket{\psi^{\prime}_k}\bra{\psi^{\prime}_{k'}},
\end{equation}
which is diagonalized to obtain  
the eigenenergies $\{E^{\prime}_k\}_{1 \leq k \leq K}$ in the ascending order and 
the corresponding eigenstates $\{ \ket{\Psi^{~}_k} \}_{1 \leq k \leq K}$, i.e., 
\begin{equation}
\hat{H}^{\prime}_{\rm tot} \ket{\Psi^{~}_k} = E^{\prime}_k \ket{\Psi^{~}_k}.
\label{eq:hp}
\end{equation}

For the next iteration, 
as the initial classical states $\ket{i}$ in Eq.~(\ref{eq:hprod}), 
we select $M\,(<K)$ classical states for approximating the $M'\,( \leq M)$ lowest eigenstates 
$\{ \ket{\Psi^{~}_{i'}} \}^{~}_{1\leq i' \leq M'}$; if we focus only on the ground state, we adopt $M'=1$. 
The selection of the states is made on the basis of the following quantity: 
\begin{equation}
p^{~}_{M',k} := \bra{\psi^{\prime}_k} \left( \frac{1}{M^{\prime}} \sum_{i'=1}^{M'} \ket{\Psi^{~}_{i'}} \bra{\Psi^{~}_{i'}} \right) \ket{\psi^{\prime}_k}.
\end{equation}
Namely, we select $M$ classical states $\ket{\psi^{\prime}_k}$ that represent $M$ largest values among 
$\{ p^{~}_{M',k} \}_{1 \leq k \leq K}$. 
Using these classical states as the initial states $\ket{i}$ in Eq.~(\ref{eq:hprod}), we repeat the above procedure 
until the $M$ lowest eigenenergies $\{E^{\prime}_k\}_{1 \leq k \leq M}$ in Eq.~(\ref{eq:hp}) converge within the desired accuracy.

Assuming that $\lambda \ll 1$ and the classical states $\ket{\psi^{\prime}_{k}}$ are not degenerate, one can simply use 
the lowest-order perturbation theory to approximately solve the eigenvalue problem in Eq.~(\ref{eq:hp}), i.e., 
\begin{equation}
\ket{\Psi^{~}_k} = \ket{\psi^{\prime}_{k}} + \lambda\sum_{k'\,(\ne k)}^K
\ket{\psi^{\prime}_{k'}} \frac{\bra{\psi^{\prime}_{k'}} \hat{H}_1 \ket{\psi^{\prime}_k}} 
{E^{~}_{\psi^\prime_k} - E^{~}_{\psi^\prime_{k'}}} 
\end{equation}
and
\begin{equation}
 E^{\prime}_k =  E_k + \lambda^2\sum_{k'\,(\ne k)}^K \frac{ | \bra{\psi^{\prime}_{k'}} \hat{H}_1 \ket{\psi^{\prime}_k} |^2 }
 {E^{~}_{\psi^\prime_k} - E^{~}_{\psi^\prime_{k'}}}. 
\end{equation}
Therefore, in the limit of the infinitesimal perturbation, the selection of the $M$ classical states based on $p^{~}_{M',k}$ is 
identical with the selection of the $M$ classical states having the $M$ lowest eigenenergies $\{ E^{~}_{\psi'_k} \}_{1 \leq k \leq M}$,  
implying that the evaluation of $\hat{H}'_{\rm tot}$ and its diagonalization are not required. 
This argument can be generalized even when there exist degenerate classical states. 
Hence, in the classical limit of the Krylov subspace method, 
we can 
search the low-energy states of $\hat{H}_0$. This is the main physical idea of the QI search method and 
more details are described in the next section. 
The success probability of the QI search method can 
be improved by increasing $M$ 
or
by optimizing the form of the quantum Hamiltonian $\hat{H}_1$. 
Here, 
we choose as $\hat{H}_1$ one- and two-body spin flipping terms, namely, 
$\hat{\sigma}^{x}_{i}$ and $\hat{\sigma}^{x}_{i}\hat{\sigma}^{x}_{j}$ for all possible sites $i$ and pairs of sites $i$ and $j$ with 
$i \neq j$, respectively.

\section{Algorithm}
\label{sec:algorithm}

In this section, we describe algorithms for the QI search method in detail.
First, we introduce 
$\ket{\sigma} \in \{ \ket{0}, \ket{1} \}$ as a complete set of a local spin state, which are eigenstates 
of the Pauli-$Z$ operator, i.e., $\hat{\sigma}^z\ket{0}=\ket{0}$ and $\hat{\sigma}^z\ket{1}=-\ket{1}$. 
Using this local basis set, we can express any state for the $N$-spin system as 
$ \ket{{\bm \sigma}=(\sigma_{0},\cdots,\sigma_{N-1})} $, which forms the complete set of states. 
For the fast simulation in the classical computer, 
we employ a typical form of the state list that makes a one-to-one correspondence 
between the vector ${\bm \sigma}$ and an integer $\psi$ representing a state as follows: 
\begin{equation}
\ket{\psi} \equiv 
\ket{{\bm \sigma}}{\rm ~with~} \psi=(\sigma_{N-1} \cdots \sigma_{1} \sigma_0 )_2=\sum^{N-1}_{i=0} 2^{i} \sigma_i ~.
\label{eq:state}
\end{equation}

Next, as an basic operation in the search algorithm, 
we define a procedure to obtain the $\ell$-site spin-flipped state 
$
\ket{\psi'}=\prod_{k=1}^{\ell} \hat{\sigma}^x_{i_k} \ket{\psi}
$
for a given state $|\psi\rangle$ with $1 \leq \ell \leq N$ and $0 \leq i_1 < \cdots < i_\ell < N$. 
Here, $\hat{\sigma}^x_i$ is the Pauli-$X$ operator acting a spin at the $i$th site. 
For this operation, we employ the bitwise exclusive or operation ({\sc Bitxor}), as shown in Algorithm~\ref{algorithm:spin_flip}.
\begin{figure}[htb]
\begin{algorithm}[H]
  \caption{Multiple spin flips for an $N$-site system}
  \label{algorithm:spin_flip}
   \begin{algorithmic}[1]
    \Require{integer $\psi$ in Eq.~(\ref{eq:state}), $\ell$ with $1 \leq \ell \leq N$, and ${\bm i}=( i_1,\cdots,i_\ell )$ with $0 \leq i_1 < \cdots < i_\ell < N$.}
    \Ensure{integer $\psi'$ with $0 \leq \psi' < 2^N$.}
    \Function{Multiple\_spin\_flip}{$\psi,{\bm i},\ell$}    	
       \State $x:=\sum_{k=1}^{\ell} 2^{i_{k}}$
	   \State $\psi':=${\sc Bitxor}$(x,\psi)$
    \EndFunction
   \end{algorithmic}
\end{algorithm}
\end{figure}

The core function for the update of the state lists in the search algorithm is given 
in Algorithm~\ref{algorithm:update}. 
\begin{figure}[htb]
\begin{algorithm}[H]
  \caption{Update state lists}
  \label{algorithm:update}
   \begin{algorithmic}[1]
    \Require{integer $\psi$,  
    $\ell$,  
    ${\bm i}$,  
    $K$ with $1 \leq K < 2^N$, 
    ${\bm \psi}=\{\psi_i \}_{1\leq i \leq K+1}$, and ${\bm \psi}'=\{\psi_i' \}_{1\leq i \leq K+1}$ with $0 \leq \psi'_1 < \cdots < \psi'_{K+1} < 2^N$; 
    real ${\bm E}=\{ \langle \psi_i | \hat{H}_0 | \psi_i\rangle \}_{1\leq i \leq K+1}$} with  $E_1 \leq \cdots \leq E_{K+1}$;
    logical variable ${\bm l}=\{ l_i \}_{1 \leq i \leq K+1}$
    \Ensure{integer ${\bm \psi}$ and ${\bm \psi'}$; real ${\bm E}$; logical variable ${\bm l}$}
	\Function{update$\_$list}{$\psi,{\bm i},\ell,{\bm \psi},{\bm \psi'}, {\bm E}, {\bm l}, K$}
	\State $\psi':=${\sc Multiple\_spin\_flip}$(\psi,{\bm i},\ell)$
	\State $(p,f):=${\sc Binary\_search}$(\psi',{\bm \psi}',K)$
	\Statex \Comment{The function ``{\sc Binary\_search}$(x,{\bm x},K)$'' with an integer/real number $x$ and a ($K+1$)-dimensional ascending-ordered integer/real vector ${\bm x}$ returns an integer $p$ for $x_p \leq x < x_{p+1}$ and a logical variable $f$ which becomes True when $
	x_p = x$.}
	\If {$f={\rm False}$}
		\State $E:=\langle \psi' | \hat{H}_0 | \psi' \rangle$
		\State $({\bm \psi},{\bm \psi'},{\bm E},{\bm l}):=${\sc update\_list\_sub}($\psi', E,{\rm False},p,$
		\Statex \hspace{5cm} ${\bm \psi},{\bm \psi'}, {\bm E}, {\bm l}, K$)
	\EndIf
	\EndFunction
	\Statex
	\Function{update\_list\_sub}{$\psi', E,l,p',{\bm \psi},{\bm \psi'}, {\bm E}, {\bm l}, K$}
	\If{$E < E_{K+1}$} 
		\State $(p,f):=${\sc Binary\_search}$(\psi_{K+1},{\bm \psi}',K)$				
		\If{$p>p'$} 
		\State ${\bm \psi}'=(\psi'_1,\cdots,\psi'_{p'},\psi',\psi'_{p'+1},\cdots,$
		\Statex \hspace{4cm} $\psi'_{p-1},\psi'_{p+1} \cdots,\psi'_{K+1})$
		\Else
		\State ${\bm \psi}'=(\psi'_1,\cdots,\psi'_{p-1},\psi'_{p+1},\cdots,$
		\Statex \hspace{4cm} $\psi'_{p'},\psi', \psi'_{p'+1} \cdots,\psi'_{K+1})$
		\EndIf		
		\State $(p,f):=${\sc Binary\_search}$(E,{\bm E},K)$
		\State ${\bm E}:=(E_1,\cdots,E_p,E,E_{p+1},\cdots,E^{~}_{K})$		
		\State ${\bm \psi}:=(\psi_1,\cdots,\psi_p,\psi',\psi_{p+1},\cdots,\psi^{~}_{K})$
		\State ${\bm l}:=(l_1,\cdots,l_p,l,l_{p+1},\cdots,l^{~}_{K})$			
		\EndIf
	\EndFunction
   \end{algorithmic}
\end{algorithm}
\end{figure}
Let us now assume that we have already kept the $(K+1)$-dimensional integer vector ${\bm \psi}$ and 
the $(K+1)$-dimensional real-valued vector ${\bm E}$ specifying respectively $K+1$ classical product states 
$\{ \ket{\psi_i} \}_{1 \leq i \leq K+1}$ and 
the corresponding energies $\{E_i = \langle \psi_i | \hat{H}_0 | \psi_i\rangle \}_{1 \leq i \leq K+1}$, 
where the order of integers $\psi_1, \psi_2, \cdots, \psi_{K+1}$ is 
sorted so that $E_1 \leq E_2\leq \cdots \leq E_{K+1}$. 
Let us also assume that we have already kept the $(K+1)$-dimensional logical vector ${\bm l}=\{l_i\}_{1 \leq i \leq K+1}$ 
with $l_i \in \{ {\rm True}, {\rm False} \}$ to judge whether $\psi_i$ has already been used as an input $\psi$ for 
Algorithm~\ref{algorithm:update}. 
In addition, for the binary search of a state, we also have the $(K+1)$-dimensional integer vector ${\bm \psi'}$ 
whose elements are the same as those of ${\bm \psi}$ but are sorted in ascending order. 
In Algorithm~\ref{algorithm:update}, we input a state $\ket{\psi}$ at the starting point and perform spin flips 
to generate a new state $\ket{\psi'}$. If $\ket{\psi'} \notin {\bm \psi}'$ and the energy 
$E = \langle \psi' | \hat{H}_0 | \psi' \rangle $ is smaller than $E_{K+1}$, then we update the lists 
${\bm \psi}$, ${\bm \psi}'$, ${\bm l}$, and ${\bm E}$ by discarding $\ket{\psi_{K+1}}$ and $E_{K+1}$.

Having described Algorithm~\ref{algorithm:update} for updating the state lists, 
we now provide the main part of the algorithm for the QI single search method 
in Algorithm~\ref{algorithm:qi-search}. 
\begin{figure}[htb]
\begin{algorithm}[H]
  \caption{QI single search method}
  \label{algorithm:qi-search}
   \begin{algorithmic}[1]
    \Require{integer $N>0$, $\psi$, $J>0$, $K$, $I>0$, ${\bm n}=\{ n_j \}_{1 \leq j \leq J}$ with all $n_j > 0$, and a set of $\mathcal{I}=\{ {\bf I}_j \}_{1\leq j \leq J}$ where ${\bf I}_j = \{{\bm i}^{(j)}_n\}_{1 \leq n \leq n_j}$ with $j$-dimensional ascending-ordered integer vector ${\bm i}^{(j)}_n$}
    \Ensure{ integer ${\bm \psi}$ and ${\bm \psi}'$; real ${\bm E}$; logical variable ${\bm l}$}
	\Function{QI\_search}{$N$, $\psi$, $J$, $K$, $I$, ${\bm n}$, $\mathcal{I}$}
    \Statex \Comment{$I_{\rm M}/R_{\rm M}$: the maximum integer/real number that can be expressed in a computer.}	
    \State $E := \langle \psi | \hat{H}_0 | \psi \rangle$
	\State ${\bm \psi}=\{\psi_i \}_{1\leq i \leq K+1}$; $\psi_i := \left\{ \begin{matrix} 
    \psi & i=1 \\ 
    I_{\rm M} & {\rm otherwise}
    \end{matrix} \right.$ 
	\State ${\bm E}=\{E_i \}_{1\leq i \leq K+1}$; $E_i := \left\{ \begin{matrix} 
    E & i=1 \\ 
    R_{\rm M} & {\rm otherwise}
    \end{matrix} \right.$
    \State ${\bm \psi}':={\bm \psi}$
	\State ${\bm l}=\{l_i \}_{1\leq i \leq K+1}$; $\{l_i\} := {\rm False}$ 
	\For {$i = 1$ to $I$}
		\State $f :={\rm False}$ 
		\For {$k = 1$ to $K+1$}
			\If {$l_k= {\rm False}$}
			\State $\psi := \psi_k$; $l_k := {\rm True}$; $f :={\rm True}$; {\bf Exit}
			\EndIf
		\EndFor
		\If {$f = {\rm True}$}
			\For {$j = 1$ to $J$}
				\For {$n = 1$ to $n_j$}
				\State $({\bm \psi},{\bm \psi}', {\bm E}, {\bm l}):=${\sc update\_list}$(\psi,{\bm i}^{(j)}_{n},j,$
				\Statex \hspace{4cm}${\bm \psi},{\bm \psi'}, {\bm E}, {\bm l}, K)$			
				\EndFor
			\EndFor
		\EndIf
  \EndFor
\EndFunction
\end{algorithmic}
\end{algorithm}
\end{figure}
In order to perform this algorithm, we have to prepare the set of integers $\mathcal{I}$ for specifying the spin flip operations 
considered in $\hat{H}_1$. For example, when $\hat{H}_1$ contains all patterns of $J$-spin flips up to $J=2$, 
we set the input parameters as follows: $n_1=N$, ${\bf I}_1=\{{\bm i}^{(1)}_1=(0),{\bm i}^{(1)}_2=(1),\cdots,{\bm i}^{(1)}_{n_1}=(N-1)\}$, 
$n_2=N(N-1)/2$, and ${\bf I}_2=\{{\bm i}^{(2)}_1=(0,1),{\bm i}^{(2)}_2=(0,2)\cdots,{\bm i}^{(2)}_{n_2}=(N-2,N-1)\}$. 
We also input the number $K$ of states kept, the number $I$ of iterations, and $\psi$ for specifying the initial classical state 
of the calculation.

\begin{figure}[htb]
\begin{algorithm}[H]
  \caption{QI multi search method}
  \label{algorithm:modified_qi-search}
   \begin{algorithmic}[1]
    \Require{integer $L>0$, $N$, $J$, $K$, $I$, ${\bm n}$ , and $\mathcal{I}$}
    \Ensure{integer ${\bm \psi}$; real ${\bm E}$}
	\Function{QI\_multi\_search}{$L$, $N$, $J$, $K$, $I$, ${\bm n}$, $\mathcal{I}$}
    \State $\bm{\phi}=\{\phi_i\}_{1 \leq i \leq L}$; $\phi_i :=${\sc rand\_between}$(0,2^N-1)$
    \Statex \Comment{The function {\sc rand\_between}$(a,b)$ returns a randomly selected integer uniformly distributed in the range $[a,b]$.}
    \State $({\bm \psi},{\bm \psi}',{\bm E},{\bm l})$:={\sc QI\_search}($N$, $\phi_1$, $J$, $K$, $I$, ${\bm n}$, $\mathcal{I}$)
    \For {$i = 2$ to $L$}
        \State $({\bm \psi}_0,{\bm \psi}'_0,{\bm E}_0,{\bm l}_0)$:={\sc QI\_search}($N$, $\phi_i$, $J$, $K$, $I$, ${\bm n}$, $\mathcal{I}$)
         \For {$j = 1$ to $K+1$}
		\State $(p,f):=${\sc Binary\_search}$(\psi'_{0,j},{\bm \psi}',K)$
		\If {$f={\rm False}$}
 	        \State $({\bm \psi},{\bm \psi'},{\bm E},{\bm l}):=${\sc update\_list\_sub}($\psi'_{0,j}, E_{0,j},$
 	        \Statex \hspace{5cm} $l_{0,j},p,{\bm \psi},{\bm \psi'}, {\bm E}, {\bm l}, K$)
 	        \EndIf
 	   \EndFor
    \EndFor
\EndFunction
\end{algorithmic}
\end{algorithm}
\end{figure}
As we shall show in the next section, the QI single search method is accidentally trapped in a local minimum, depending on 
the combination of the initial state $\psi$ and the random coupling realization in $\hat{H}_0$.
One of the simplest solutions to avoid trapping in a local minimum is to perform the independent QI single searches in parallel 
for different random initial states and merge the lists obtained from the independent searches. 
The QI multi search method given in Algorithm~\ref{algorithm:modified_qi-search} is based on this strategy.

Let us now consider the computational complexity of the algorithm in Algorithm~\ref{algorithm:modified_qi-search}. 
The total number of the spin flip operations required per an iteration is $\sum_{j}n_j \sim O(N^J)$. 
For a spin-flipped classical state, we have to evaluate the energy of $\hat{H}_0$, 
which costs naively $O(N^2)$. 
However, since we know the energy of the state before the spin flip operation, the computational cost for evaluating the energy 
of the state after the spin flip operation is reduced to $O(N)$ because we only take into account the couplings 
associated with the spin-flipped sites. 
The computational cost for searching a state via the binary search is $\log_2 K$ and thus it is negligible because we consider  
the case for $K \ll 2^N$. 
Therefore, the leading order of the computational cost per an iteration with the system size $N$ is $O(N^{J+1})$.
The total computational cost in Algorithm~\ref{algorithm:modified_qi-search} is thus $O(LIN^{J+1})$. 
In this study, we set the parameters as $(K,I,L)=(K_0,K_0,1)$ and $(K,I,L)=(K_0,N,N)$ with $K_0=N(N+1)/2+1$, 
and hence the total computational cost in both cases is $O(N^{5})$, assuming that $J=2$.

\section{Basic Features of the QI search method}
\label{sec:results}
In this section, we test the method by trying to find the 120 lowest-energy states 
of the classical Ising model described by the Hamiltonian $\hat{H}_0$ in Eq.~(\ref{eq:H0}) 
with the fully connected random interactions $J_{ij}$ and the random magnetic fields $h_i$,
which are both drawn from a uniform distribution on the interval $[-1/2,1/2]$. 
We consider 120 different instances of $\hat{H}_0$ for each system size $N$ 
to quantitatively estimate accuracy and efficiency of the QI search method.
The results for $N$ up to 30 sites are compared with those obtained 
exactly by the brute-force search.
For larger systems with $N\sim O(10^3)$, 
the results are compared with those obtained by
the PTMC method~\cite{Hukushima1996,Hukushima_IntJModPhys1996},
focusing only on the ground-state energy.

\subsection{Numerical setup and elapsed time}

We first investigate the elapsed time of the QI search method given in Algorithm~\ref{algorithm:modified_qi-search} 
with the following input parameters: $L=1$ (i.e., equivalent to the QI single search method given in Algorithm~\ref{algorithm:qi-search}), 
$\ket{\phi_1=0} \equiv \ket{00 \cdots 0}$ for the initial classical state, and $K=I=\sum_{j=1}^2n_j+1$ with 
$\sum_{j=1}^2n_j=N+N(N-1)/2=N(N+1)/2$, where we assume the infinitesimal quantum Hamiltonian
\begin{equation}
\hat{H}_1 = \epsilon \left( \sum_{i<j} \hat{\sigma}^x_i \hat{\sigma}^x_j + \sum_{i} \hat{\sigma}^x_i \right)
\label{eq:H1} 
\end{equation}
with 
an infinitesimally small real number $\epsilon$. 
Note that the one- and two-spin flip operations considered in $\hat{H}_1$ 
correspond to the example for ${\bf I}_1$ and ${\bf I}_2$, respectively, given in Sec.~\ref{sec:algorithm}.

\begin{figure}[htb]
\begin{center}
\includegraphics[width=8cm]{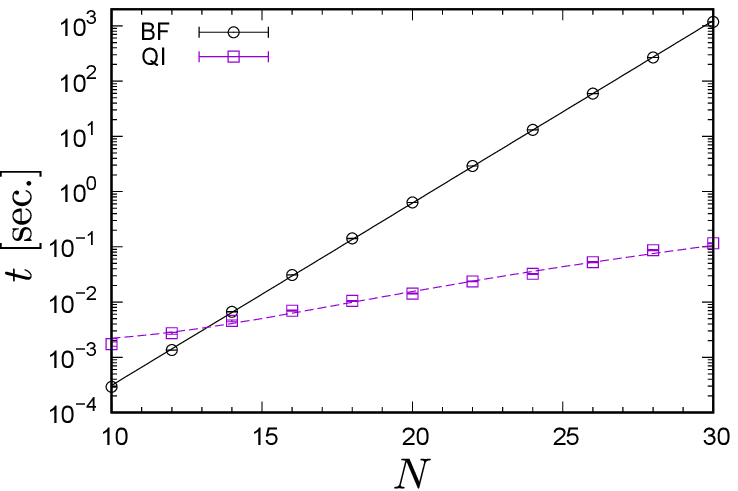}
\caption{ 
Averaged elapsed time $t$ for the brute-force (BF) search and the QI search with 
$L=1$, $K=I=N(N+1)/2+1$, $J=2,n_1=N$, and $n_2=N(N-1)/2$ as a function of the system size $N$. 
The black solid line and the purple dashed curve are fitting functions $\ln t = a_1N + b_1$ with $(a_1,b_1) = (0.76,-15.7)$ and 
$t=a_2N^5 + b_2$ with $(a_2,b_2) = (4.7\times 10^{-9},1.8\times 10^{-3})$, respectively.
}
\label{fig:time_vs_N}
\end{center}
\end{figure}

According to the discussion toward the end of Sec.~\ref{sec:algorithm}, 
we expect that the leading order of the computational cost 
of the QI search method with these parameters is $O(N^5)$, 
which clearly has better scalability as compared to $O(2^N)$ for the brute-force search. 
To confirm the scalability, 
we perform the QI search on the HOKUSAI BigWaterfall (CPU: Intel Xeon Gold 6148 2.4GHz) 
installed at RIKEN for the same sets of Hamiltonian $\hat{H}_0$ treated in Sec.~\ref{Sec:H0} 
and measure the averaged elapsed time without any parallelization. 
As shown in Fig.~\ref{fig:time_vs_N}, we confirm that 
the elapsed time $t$ of the QI search is nicely on the line of $t \sim N^5$ and 
is shorter 
than that of the brute-force search for $N \geq 14$.

\subsection{$N$, $L$, and $K$ dependence 
of the success probability 
for searching the ground state 
}

As the basic performance of the QI search method, we next investigate the $N$, $L$, and $K$ dependence of the success 
probability $p_1$ for searching the ground state of the classical Ising Hamiltonian $\hat{H}_0$. Here, $p_1$ is defined as the 
ratio between the number of instances of the random coupling realizations in $\hat{H}_0$ for which the ground state is correctly 
searched and the total number of instances of the random coupling realizations in $\hat{H}_0$, which is 120 for each system size 
$N$ in this study. 

\begin{figure}[htb]
\begin{center}
\includegraphics[width=8cm]{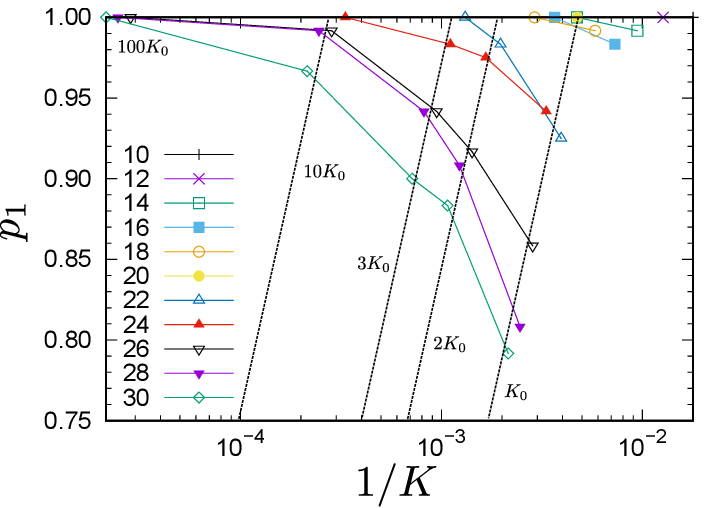}
\caption{
Success probability $p_1$ of the QI search with $(L,\phi_1)=(1,0)$ and $I=K$ 
for different system sizes $N=10,12,\cdots,30$ indicated in the figure. 
The dotted lines are guide for the eye to highlight the slow decay of the success probability $p_1$ with increasing $N$ for 
each $K\in\{K_0,2K_0,3K_0,10K_0,100K_0\}$ where $K_0=N(N+1)/2+1$.
}
\label{fig:N_K_N1}
\end{center}
\end{figure}

We first perform the QI search by setting the parameters $(L,\phi_1)=(1,0)$ and $I=K$ for the system sizes 
up to $30$ sites.  
The number of lowest-energy states kept is set to be $K=aK_0$ with $a \in \{1, 2, 3, 10, 100\}$ and $K_0=N(N+1)/2+1$. 
We can naively expect that $p_1$ decreases exponentially with increasing $N$ for a fixed value of $a$ because the number of 
classical states generated in the QI search for this setting is only $O(IN^2) \sim O(N^4)$, which is extremely 
smaller than the dimension $2^N$ of the total Hilbert space.
However, as shown in Fig.~\ref{fig:N_K_N1}, the probability $p_1$ decays rather slowly 
instead of exponentially 
for each value of $a$, and rapidly increases to close to 1 with increasing $a$. 
These characteristics strongly support the high scalability of the QI search method.

\begin{figure}[htb]
\begin{center}
\includegraphics[width=8cm]{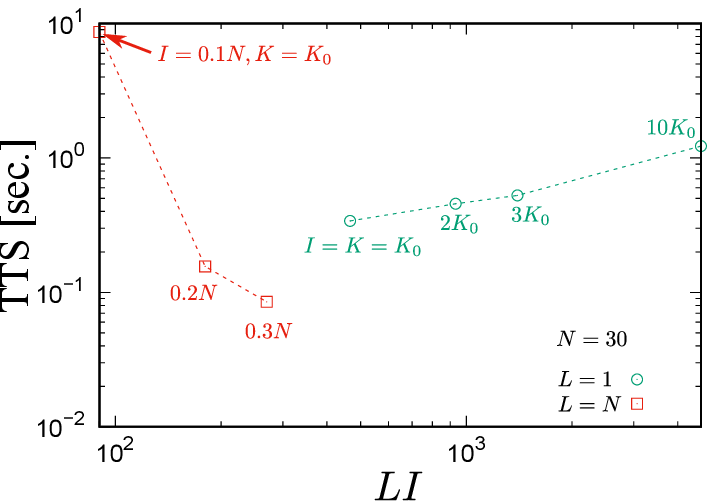}
\caption{
Time to solution (TTS) versus the total number of QI iterations, $LI$, for searching the ground state of 120 different 
instances of the random coupling realizations in $\hat{H}_0$ with $N=30$. 
The QI multi search method with $L=N$, 
$K=K_0$, and $I\in\{0.1N,0.2N,0.3N,0.5N\}$ is employed (red squares). For comparison, the results for the QI single search method 
with $L=1$, $I=K$, $K\in\{K_0,2K_0,3K_0,10K_0,100K_0\}$, and $K_0=N(N+1)/2+1$ are also shown by green circles 
(these results are the same as those shown in Fig.~\ref{fig:N_K_N1}). 
Note that the computational complexity for these two methods with these parameters is $O(N^5)$. 
The results for $(I,K,L)=(0.5N,K_0,N)$ and $(I,K,L)=(100K_0,100K_0,1)$ are not plotted 
because the TTS is zero. 
}
\label{fig:itr_tot_vs_tts}
\end{center}
\end{figure}

Second, we discuss the efficiency of the QI multi search with $L>1$ 
by setting the input parameters $L=N=30$, $K=K_0$, and $I\in\{0.1N,0.2N,0.3N,0.5N\}$, where  
the 30 initial classical states are randomly chosen. 
To quantify the accuracy and efficiency of the method, 
here we estimate a time-to-solution (TTS) that measures the time required 
to obtain the correct ground state 
with a successful probability of 0.99~\cite{Ronnow2014,Aramon2019,Rozada2019}. 
The TTS is defined as 
\begin{equation}
{\rm TTS}=\frac{\ln(1-0.99)}{\ln(1-p^{~}_1)} t, 
\label{eq:tts}
\end{equation}
where $t$ is the elapse time of the QI search 
and we use the averaged elapsed time over 120 instances of the random coupling realizations. 
Figure~\ref{fig:itr_tot_vs_tts} shows the results of the TTS for the QI multi search. 
For comparison, the results for the QI single search with $N=30$ (the same results shown in Fig.~\ref{fig:N_K_N1}) 
are also plotted. 
Note that the QI single search in Fig.~\ref{fig:N_K_N1} sets $L=1$ but $I \sim O(N^2)$, while  
the QI multi search here sets $L=N$ but $I \sim O(N)$. 
Hence, the total computational cost is $O(N^5)$ for both cases. 
Note also that the ground states for all the 120 different instances 
of the random coupling realizations in $\hat{H}_0$ 
are found correctly 
by the QI search with $(I,K,L)=(0.5N,K_0,N)$ and $(I,K,L)=(100K_0,100K_0,1)$, 
for which the TTS is zero, and thus these results are not plotted in Fig.~\ref{fig:itr_tot_vs_tts}. 
As shown in Fig.~\ref{fig:itr_tot_vs_tts}, the QI multi search is already better that the QI single search 
in this case because the TTS is smaller, provided that the effective total number of 
iterations, i.e., $LI$, is the same. 
Considering that parallelization of the QI search for different initial states with $L=N$
yields $O(N)$ speed-up, we expect that the TTS for the QI multi search becomes much smaller, 
which is obviously a feature advantageous for using 
supercomputers 
when the problem size is particularly large.

\begin{figure}[htb]
\begin{center}
\includegraphics[width=8cm]{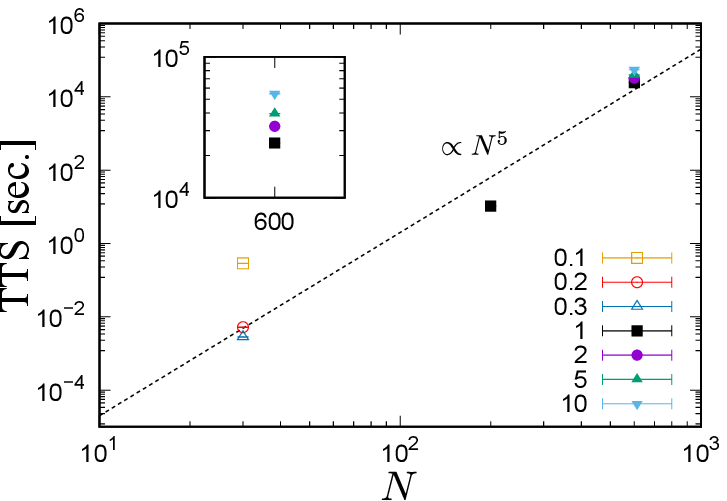}
\caption{ 
System-size $N$ dependence of the time to solution (TTS) for searching the ground state of 120 different 
instances of the random coupling realizations in $\hat{H}_0$ with $N=30, 200, 600$. 
Numbers in the legend indicate the value of $a$ for the input parameters $(L,K,I)=(N,aN,aN)$. 
Inset: enlarged view for the data of $N=600$. 
Since the brute-force search cannot be applied for $N=200$ and 600, 
the PTMC method is used to obtain the reference ground-state energy for these large systems 
(see Appendix~\ref{sec:estimation_quality}). 
}
\label{fig:tts_vs_N}
\end{center}
\end{figure}

To confirm the scalability of the QI search method, 
Fig.~\ref{fig:tts_vs_N} shows the results of the TTS for the QI multi search with 
OpenMP parallelization in terms of $L$. 
Here we adopt several different input parameters $(L, K, I)$ for the QI search and 
employ the PTMC method to obtain the reference ground-state energy for large systems.
As shown in Fig.~\ref{fig:tts_vs_N}, we find that the TTS scales approximately as $N^5$, which is comparable with 
the scaling of the computational cost $O(N^{5})$. 
The numerical details for larger systems and the comparison with the PTMC results are 
described in Appendix~\ref{sec:estimation_quality}.

\subsection{Instance and initial state dependence 
of the number of iterations required for searching the ground state 
}

In order to understand the reason why the QI multi search method with $L>1$ improves the success probability $p_1$, 
we now focus on the instance and initial state dependence of the number $I_1$ of iterations necessary to find the correct 
ground state of $\hat{H}_0$. 

\begin{figure}[htb]
\begin{center}
\includegraphics[width=8cm]{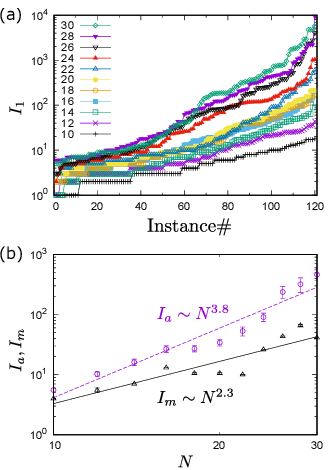}
\caption{ 
(a) Instance dependence of the number $I_1$ of iterations required to find the correct ground state of the Hamiltonian 
$\hat{H}_0$ for 120 different instances of the random coupling realizations with the system size $N=10,12, \cdots,30$ indicated 
in the figure. Note that the instance number is ordered so that $I_1$ monotonically increases for each $N$. 
(b) Ensemble-averaged $I_1$ (named $I_a$) and the median of the ensemble-dependent $I_1$ (named $I_m$) as a function 
of the system size $N$. The solid and dashed lines are power law fittings of $I_m$ and $I_a$, respectively. 
}
\label{fig:I1_Ia_Im}
\end{center}
\end{figure}

For this purpose, we first employ the QI single search method with the input parameters $(L,\phi_1)=(1,0)$ and $I=K$ for 
$K$ up to $K=100K_0$. 
Figure~\ref{fig:I1_Ia_Im}(a) shows the results of $I_1$ for 120 different instances
of the random coupling  realizations in $\hat{H}_0$ with different system sizes.  
In this figure, the instance number is reordered so that $I_1$ monotonically increases. 
We find that $I_1$ basically increases with the system size $N$. 
To capture a feature of this increase, we also evaluate the ensemble-averaged $I_1$, named $I_a$, 
and the median of the ensemble-dependent $I_1$, named $I_m$, as a function of $N$. 
As shown in Fig.~\ref{fig:I1_Ia_Im}(b), we find that $I_a > I_m$ for all the system sizes. 
This is simply because $I_1$ around instance \#120 shown in 
Fig.~\ref{fig:I1_Ia_Im}(a) is especially large as compared to that for other instances. 
In accordance with the results in Fig.~\ref{fig:N_K_N1}, 
$I_a$ and $I_m$ do not seem to increase exponentially with $N$. 
Therefore, as shown in Fig.~\ref{fig:I1_Ia_Im}(b), we fit $I_a$ and $I_m$ with power law functions and find that the exponent of $N$ for 
$I_m$ is smaller than that for $I_a$. 
This fitting implies that the ground states of the half of the instances 
(i.e., the Hamiltonian $\hat{H}_0$ with different random coupling realizations) 
for the system size $N \leq 30$ 
can be searched successfully with the number of iterations up to only $I_1 \sim O(N^{2.3})$.
Having obtained the different exponents, we speculate that the QI single search for the Hamiltonian 
$\hat{H}_0$ with the random coupling realizations around sample \#120 is trapped in a local minimum since here we employ only 
a single initial state with $\ket{\phi_1=0} \equiv \ket{00 \cdots 0}$. 

\begin{figure}[htb]
\begin{center}
\includegraphics[width=8cm]{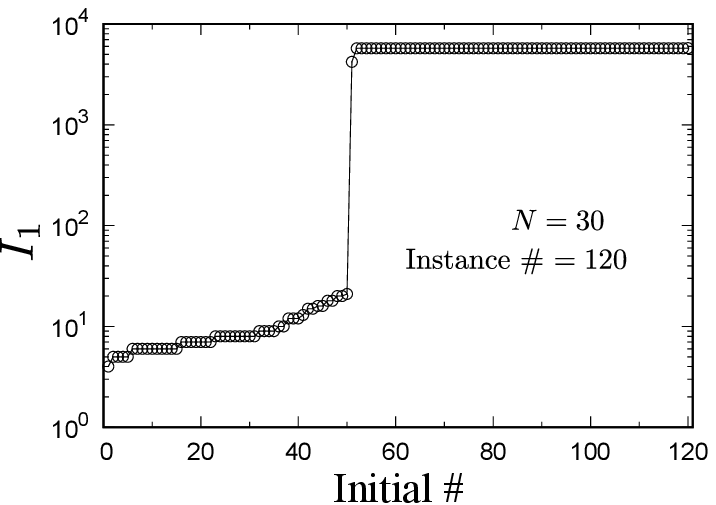}
\caption{
Initial state dependence of the number $I_1$ of iterations. The Hamiltonian $\hat{H}_0$ is selected for the random coupling realization of 
instance \# 120 in Fig.~\ref{fig:I1_Ia_Im}(a) with the system size $N=30$. 
The QI single search method is employed with $120$ different initial states selected randomly and 
the initial number for specifying the different initial states is ordered so that $I_1$ increases monotonically. 
}
\label{fig:I_N_random_init}
\end{center}
\end{figure}

To confirm this assertion, we next explore the initial state dependence of $I_1$ by performing 
the QI single search 
using $120$ different initial states for the Hamiltonian $\hat{H}_0$ of instance \#120 in Fig.~\ref{fig:I1_Ia_Im}(a). 
Figure~\ref{fig:I_N_random_init} shows the results of $I_1$ as a function of the initial number for specifying the $120$ different 
initial states that is ordered so that $I_1$ increases monotonically.  
We find that 40\% of the 120 different QI searches can find the ground state successfully with only $I_1<N$ iterations.
This observation clearly supports that the QI multi search with $L>1$ is an efficient strategy to avoid trapping local minima.

\subsection{Search efficiency for low-energy excited states}

\begin{figure}[htb]
\begin{center}
\includegraphics[width=8cm]{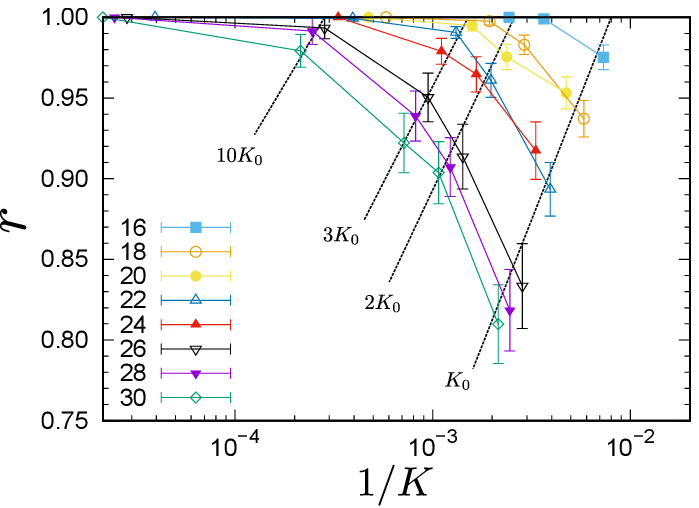}
\caption{ 
Success rate $r$ of the number of low-energy states successfully detected among the 120 lowest-energy states 
for a given random coupling realization, which is averaged 
over 120 different instances of the random coupling realizations in $\hat{H}_0$, 
with different system sizes $N=16,18,\cdots,30$ 
indicated in the figure. 
The error bars indicate the standard error. 
The QI single search is employed with the input parameters $(L,\phi_1)=(1,0)$ and $I=K$. 
The dotted lines are guide for the eye to highlight the slow decay of the success rate $r$ with increasing $N$ for 
each $K\in\{K_0,2K_0,3K_0,10K_0,100K_0\}$, where $K_0=N(N+1)/2+1$.
}
\label{fig:nc_av_std}
\end{center}
\end{figure}

In this section,
we shall investigate the search efficiency of the QI search for the low-energy excited states.
Note that when we perform the QI single search for obtaining the results shown in Fig.~\ref{fig:N_K_N1}, 
$K$ lowest-energy states, in addition to the ground state, are also obtained. 
This is an advantage of the QI search method proposed in this study. 
Here, aiming to find $120$ lowest-energy states, we evaluate the ensemble-averaged success 
rate
$r$  
between the number of states correctly found and the number of the target states, namely, 120.
Figure~\ref{fig:nc_av_std} shows the results obtained by the QI single search 
with the input parameters $(L,\phi_1)=(1,0)$, $I=K=aK_0$, 
and $a\in\{1,2,3,10,100\}$ for $N$ up to 30 sites. 
Surprisingly, 
the $K$ dependence of $r$ is very similar to that of the success probability $p_1$ 
for searching the ground state shown in Fig.~\ref{fig:N_K_N1}. 
This implies that the QI single search method given in Algorithm~\ref{algorithm:qi-search} 
is highly scalable not only for the ground state 
but also for the low-energy excited states.

\begin{figure}[htb]
\begin{center}
\includegraphics[width=8cm]{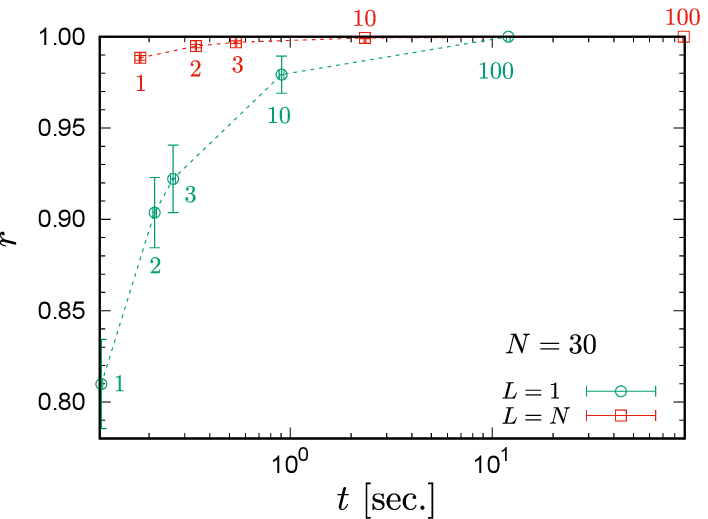}
\caption{ 
Success rate $r$ versus the elapsed time $t$ for searching the 120 lowest-energy states of $\hat{H}_0$ with the system size 
$N=30$. $r$ and $t$ are averaged over 120 different instances of the random coupling realizations in $\hat{H}_0$. 
The error bars indicate the standard error.
The QI single search method with $(L,K,I)=(1,aK_0,aK_0)$ (denoted by green circles)
and the QI multi search method with $(L,K,I)=(N,aK_0,aN)$ (denoted by red squares)
are employed. Here, $a\in\{1,2,3,10,100\}$ and is indicated beside each symbol. 
}
\label{fig:r_vs_t}
\end{center}
\end{figure}

Figure~\ref{fig:r_vs_t} shows the success rate $r$ versus the averaged elapsed time $t$ 
for the QI single search with $(L,K,I)=(1,aK_0,aK_0)$ and 
for the QI multi search with $(L,K,I)=(N,aK_0,aN)$. The system size is $N=30$ for both cases.
Note that the overall computational costs for these two methods with these parameters 
both scale as $N^5$. 
We find in Fig.~\ref{fig:r_vs_t} that 
the success rate $r$ for the QI multi search with $L=N$ is always better 
than that for the QI single search with $L=1$, 
comparing both data for a common averaged elapsed time $t$.
For example, comparing the result with $a=2$ for the QI single search and that with $a=1$ in the QI multi search, 
which cost comparable elapsed time $t\sim 0.2$ seconds, the error $\epsilon=1-r$ of the search in the latter is about 8.3 times 
smaller than that in the former. 
We should also emphasize that 
the QI multi search with $I=N$ can already find about 99\% of the 120 lowest-energy states, and 
the success rate $r$ becomes even better with larger $I$. 
Thus, the QI multi search with $L>1$ can efficiently search not only the ground state but also 
the low-energy excited states.

\section{Basic physical property of $\hat{H}_0$}
\label{Sec:H0}

Finally, in this section, we briefly study 
the basic physical properties of the random Ising model described by the Hamiltonian $\hat{H}_0$ in Eq.~(\ref{eq:H0}), 
such as the ground-state and first-excited energies and the number of states in the low-energy region. 
For this purpose, we employ the QI search method as well as the brute-force method and the PTMC method
for the system sizes up to $N=1000$. 
Note that the Hamiltonian $\hat{H}_0$ with the couplings ($J_{ij},h_i)$ replaced 
with $(J_{ij}/\sqrt{N},0)$ and $J_{ij}$ being distributed according to the Gaussian distribution 
is well known as the Sherrington-Kirkpatric (SK) model~\cite{SK1975}, and the ground state 
of the SK model has been studied by various numerical methods such as 
genetic algorithms~\cite{Palassini1999,Palassini2008}, 
hierarchical methods~\cite{Bouchaud2003}, 
extremal optimizations~\cite{Boettcher2005}, 
and conformational space annealing~\cite{Kim2007}.

\begin{figure}[htb]
\begin{center}
\includegraphics[width=8cm]{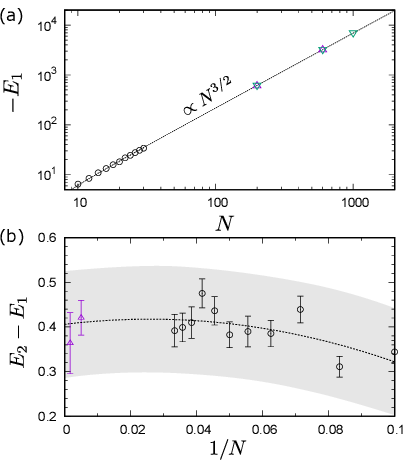}
\caption{
(a) Ensemble-averaged ground-state energy $E_1$ as a function of the system size $N$. 
The results are obtained by the brute-force method (circles), the QI search method (upper triangles), and the PTMC method 
(lower triangle). 
Broken line is the fitting function $E_1= aN^{3/2} + b$ with $a=-0.218\pm0.001$ and $b=0.79\pm0.14$. 
(b) Ensemble-averaged first excitation gap $E_2-E_1$ as a function of $1/N$ obtained by the brute-force method (circles) 
and the QI search method (upper triangles). 
The error bars indicate the standard error. 
The dashed line is the quadratic-fit for the gap and the gray shade indicates the fitting error in the thermodynamics limit at $1/N\to0$. 
In the cases of $N=200$ and $600$, denoted by upper triangles in (a) and (b), 
we only use the results of the random coupling instances for which 
the QI search method and the PTMC method give exactly the same ground-state energy (see Appendix~\ref{sec:estimation_quality}).
}
\label{fig:e0_and_gap}
\end{center}
\end{figure}

To evaluate the physical quantities, 
we consider 120 instances of the random coupling realizations in $\hat{H}_0$ 
for each system size $N$. As in Sec.~\ref{sec:results}, all the coupling $J_{ij}$ and $h_i$ in $\hat{H}_0$ are 
chosen randomly from a uniform distribution on the interval $[-1/2,1/2]$. 
Figure~\ref{fig:e0_and_gap}(a) reveals that the ensemble-averaged ground-state energy $E_1$ 
is clearly proportional to $N^{3/2}$ and is well fitted by a function $E_1= aN^{3/2} + b$ 
with $a=-0.218\pm0.001$ and $b=0.79\pm0.14$.
The leading exponent $N^{3/2}$ is exactly the same as that of the SK model~\cite{SK1975}, 
taking into account the $N$ dependence of the random couplings in the SK model, i.e., $J_{ij}/\sqrt{N}$.
In Fig.~\ref{fig:e0_and_gap}(b), we also estimate the first excitation gap in the thermodynamic limit by a quadratic fit of 
the ensemble-averaged first-excited energy $E_2-E_1$ with respect to $1/N$,
and find that there exists a finite gap in the thermodynamic limit as large as $E_2-E_1=0.41\pm0.12$. 

\begin{figure}[htb]
\begin{center}
\includegraphics[width=8cm]{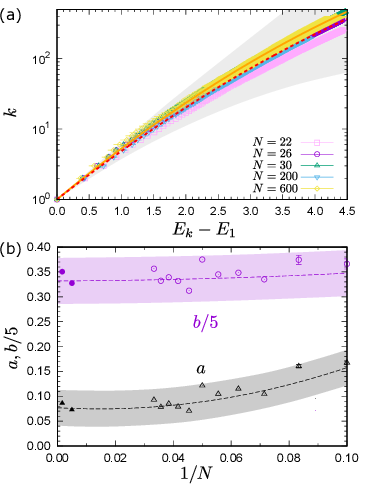}
\caption{ 
(a) Ensemble-averaged energy difference $E_k-E_1$ between the $k$th lowest energy $E_k$ 
and the ground-state energy $E_1$ with different system sizes up to $N=600$ (see the text for details of the ensemble). 
The orange line 
is the fitting function $\ln k = -a (E_k-E_1)^2 + b (E_k-E_1)$ with $a=0.0927\pm0.0003$ and $b=1.783\pm0.002$ 
obtained for the results of $N=30$. 
The red dashed line is the fitting function in the thermodynamic limit with $a=0.077\pm0.036$ and $b=1.66\pm0.23$ 
estimated in (b) 
and the gray shade indicates the fitting error. 
(b) Fitting parameters $a$ and $b$ as a function of $1/N$. 
These parameters are obtained by fitting the results of $k$ vs. $E_k-E_1$ for each system size $N$, as in (a). 
The purple and black dashed lines are the quadratic fits for these parameters $a$ and $b$, respectively. 
The purple and gray shades indicate the fitting errors of $a$ and $b$, respectively, in the thermodynamic limit at $1/N\to0$. 
}
\label{fig:k_vs_Ek}
\end{center}
\end{figure}

We also investigate the total number of states $k(E)$ below a certain energy $E$ in the low-energy region.
This quantity can be useful, 
since the first derivative of $k(E)$ with respect to $E$ is related to the density of states.
We simply obtain the $k$th lowest energy $E_k$, which is averaged over the 120 different instances 
for $N \leq 30$.
For the larger systems with $N=200$ and $600$, 
we assume that the low-energy states are correctly obtained by the QI search method in the cases 
where the QI search method and the PTMC method estimate the same ground-state energy. 
Such cases are found for 95 (26) random coupling instances with $N=200~(600)$, as described in Appendix~\ref{sec:estimation_quality}, 
and only these results are used for the ensemble average. 
Figure~\ref{fig:k_vs_Ek}(a) shows the semi-log plot of $k$ vs. $E_k-E_1$ in the low-energy region 
for different system sizes up to $N=600$. 
We find that $\ln k$ is almost proportional to $E_k-E_1$ in the vicinity of the ground state
$E_k-E_1 \sim 0$ and gradually shifts to an upward-convex function with increasing 
$k$.
As a phenomenological fitting function, 
we propose $\ln k = -a (E_k-E_1)^2 + b (E_k-E_1)$ with positive fitting parameters $a$ and $b$. 
Applying the least-square fitting to the date for $N=30$, 
we obtain the fitting curve with $a=0.0927(3)$ and $b=1.783(2)$ that is indicated by the orange 
line in Fig.~\ref{fig:k_vs_Ek}(a), confirming that the fitting curve well fits the data 
for $E_k-E_1 \leq 4.5$. 

Next, we estimate the fitting curve in the thermodynamic limit
by first fitting the data of $k$ vs. $E_k-E_1$ for each system size $N$ to obtain $a$ and $b$.  
Then we perform quadratic fittings of $a$ and $b$ with respect to $1/N$, as shown in Fig.~\ref{fig:k_vs_Ek}(b), and estimate 
the parameters $a$ and $b$ in the thermodynamic limit 
by extrapolating them to $1/N\to0$. 
The fitting curve with $a$ and $b$ in the thermodynamic limit is indicated by the red dashed 
line in Fig.~\ref{fig:k_vs_Ek}(a),  
which is comparable with the data for $N$ up to 600 in the region of $E_k-E_1 \lesssim 1$.
Notice also that the first-excited energy $E_2-E_1$ is insensitive to $N$ and it is almost on the red dashed line, 
thus consistent with the results shown in Fig.~\ref{fig:e0_and_gap}(b).

\section{Summary and discussion}
\label{sec:summary}

We have proposed a QI search method for finding low-energy states of 
the classical Ising model described by the Hamiltonian $\hat{H}_0$ in Eq.~(\ref{eq:H0}) with 
the random Ising interactions $J_{ij}$ and the random local magnetic fields $h_i$, 
to which many discrete optimization problems can be mapped. 
The main idea of the QI search algorithm is based on a classical limit of 
the $M$-block two-step Krylov subspace method, which is known as a powerful method in the numerical 
linear algebra to calculate the low-energy eigenstates of a matrix. 
An essential point of the QI search algorithm is to introduce infinitesimal quantum interaction 
$\hat{H}_1$ that is not commutable to the original classical Hamiltonian $\hat{H}_0$ and 
expand the Krylov subspace for the total Hamiltonian $\hat{H}_0+\hat{H}_1$. 

The performance of the QI search method is analyzed for the Hamiltonian $\hat{H}_0$ 
with $\{J_{ij},h_i\}$ being distributed uniformly in the range of $[-1/2,1/2]$.  
For this purpose, we considered 120 instances of the random coupling realizations in $\hat{H}_0$.
We found that the QI multi search method given in Algorithm~\ref{algorithm:modified_qi-search}  
with $L=N$ random initial states can successfully obtain the correct ground states for all the 120 
instances with the system size $N$ up to 30 only by $I=N/2$ iterations. 
In addition, we have shown that 99\% of the 120 lowest-energy states are found correctly within $I=N$ iterations. 
The search accuracy is improved monotonically with increasing the number $K$ of states kept and the number $I$ of iterations. 
We have also found that the TTS of the QI search method with OpenMP parallelization 
for searching the ground state scales approximately as $N^5$ with the system size $N$ up to 600. 
The overall computational complexity of the QI search method is $O(LIN^{J+1})$, 
and the algorithm can be easily parallelized with respect to $L$, 
thus compatible with the calculations for large systems using supercomputers.

We have also briefly investigated the low-energy properties of the random coupling Ising model described by the Hamiltonian 
$\hat{H}_0$ using the QI search method as well as the brute-force method and the PTMC method for the system size $N$ up to 
$1000$. 
We have found that the ensemble-averaged ground-state energy $E_1$ scales as $N^{3/2}$,  
which is consistent with the SK model, 
taking into account the fact that the coupling $J_{ij}$ here is not scaled with $N$. 
We have also found that there exists a finite energy gap between the ground state and the first 
excited state in the thermodynamic limit. 
Moreover, we have found that the total number $k(E)$ of states below a certain energy $E$ in the low-energy region 
increases exponentially with the excitation energy.

It would be interesting to explore the search efficiency of the QI search 
with $\hat{H}_1$ being only the uniform magnetic field 
$\epsilon  \sum_{i} \hat{\sigma}^x_i $ with $J=1$. 
Since the computational cost with $J=1$ is only $O(N^2)$ in each iteration, 
comparable with the SA method, we can treat the system size up to $O(10^4)$.  
However, since the number of classical states generated per iteration is only $O(N)$, not $O(N^2)$ as in the case of $J=2$ 
considered here in this study, 
we naively expect that the state list is much less updated in the QI search with $J=1$, 
assuming the same number of $K$. 
As a consequence, the QI search with $J=1$ may be easily trapped in local minima. 
Nonetheless, it is worthwhile to perform the QI search with $J=1$ for the ground-state search of the system size 
as large as $O(1000)$ and compare the results with other numerical methods such as the SA and PTMC methods,  
which is left for a future study.

In the classical Ising Hamiltonian $\hat{H}_0$ in Eq.~(\ref{eq:H0}), 
the total $\hat{\sigma}^z$ is a good quantum number. 
In order to preserve this symmetry, 
we could introduce as the infinitesimal quantum interaction $\hat{H}_1$ 
the infinitesimal XX interaction 
$\epsilon\sum_{i<j}(\hat{\sigma}^x_i\hat{\sigma}^x_j + \hat{\sigma}^y_i\hat{\sigma}^y_j)$ 
by replacing the spin flip operations in Algorithm~\ref{algorithm:spin_flip} 
with the spin exchange operations, 
and search the ground state and low-energy states for each sector of the total $\sigma^z$. 
Furthermore, 
the classical Hamiltonian $\hat{H}_0$ considered in statistical physics and 
condensed-matter physics often has the translational and point group symmetries. 
We can also implement these symmetry constraints in the QI search algorithm 
by introducing a representative state for each sector with different quantum numbers 
associated with the symmetry~\cite{sandvik2010}, 
which is often used in the exact diagonalization method.

The QI search algorithm proposed here is inspired by the Krylov subspace method 
in the sense that the infinitesimal quantum interaction $\hat{H}_1$ is considered 
to generate classical products states, and those states having higher energies are
then truncated to keep the number of lower energy states manageable.
As we have demonstrated, after repeatedly applying this procedure, 
the ground state as well as the low-energy states of the classical Ising Hamiltonian $\hat{H}_0$ 
is successfully obtained. 
A similar algorithm based on the same strategy can be applied to obtain the ground state of the quantum 
Hamiltonian $\hat{H}=\hat{H}_0+\hat{V}_1$, where $\hat{H}_0$ is the classical 
Hamiltonian and $\hat{V}_1$ is the perturbatively small quantum Hamiltonian.  
It is a legitimate approximation in terms of perturbation theory to calculate 
the ground state by diagonalizing the Hamiltonian $\hat{H}$ in the Hilbert space 
spanned by a finite number of classical products states, 
which are expanded by applying $\hat{V}_1$ and are then truncated according to their 
weights contributing to the approximate ground state in the expanded Hilbert space.

\acknowledgments
We would like to thank T.~Shirakawa and M.~Ohzeki for helpful comments. 
The work was partially supported 
by KAKENHI Nos. JP17K14359, JP18K03475, JP18H01183, JH20H01849, JP21K03395, JP21H03455, and JP21H04446, 
and by JST PRESTO No. JPMJPR1911. 
This work was also supported by MEXT Q-LEAP Grant Number JPMXS0120319794. 
We are also grateful for allocating us computational resources of the HOKUSAI BigWaterfall supercomputing 
system at RIKEN.

\appendix

\section{
QI search for larger systems and comparison with PTMC calculations
}
\label{sec:estimation_quality}

For a practical use, it is important to verify that the QI search method can efficiently find the ground state in large systems 
up to $O(10^3)$. 
To this end, we need the reference data for the ground-state energy in large systems that cannot be treated by the brute-force 
method. Here, we employ the PTMC method~\cite{Swendsen1986, Geyer1991,Hukushima1996,Falcioni1999,Earl2005}, 
which is well known as one of the best heuristic methods for searching the ground state of a system such as the classical Ising 
model described by the Hamiltonian $\hat{H}_0$ in Eq.~(\ref{eq:H0}).

\begin{figure}[htb]
\begin{center}
\includegraphics[width=8cm]{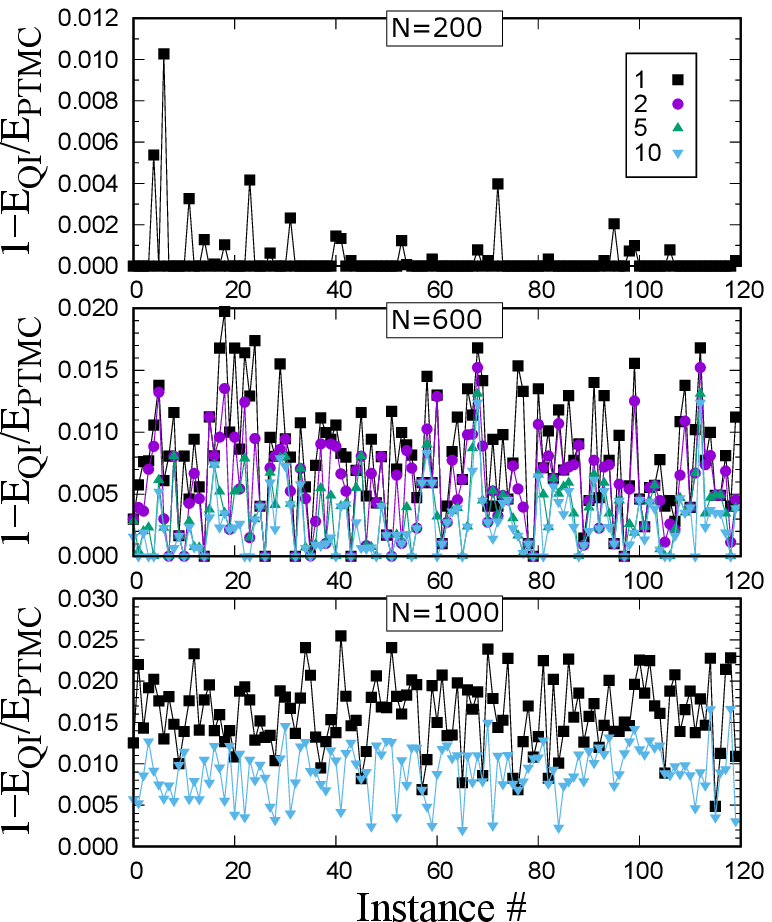}
\caption{ 
Comparison of the ground-state energies estimated by the QI search method ($E_{\rm QI}$) and the PTMC method 
($E_{\rm PTMC}$) for 
120 different instances of the random coupling realizations in $\hat{H}_0$ with $N=200$, $600$, and $1000$. 
The different input parameters $(L, K, I)=(N,aN,aN)$ with $a=1$, 2, 5, and 10 (indicated in the legend) 
are used for the QI search method. 
Note that the estimated energies $E_{\rm QI}$ and $E_{\rm PTMC}$ are both negative and thus $1-E_{\rm QI}/E_{\rm PTMC}\geq0$ 
implies $E_{\rm QI}\geq E_{\rm PTMC}$. 
}
\label{fig:ene_diff_qi_ptmc}
\end{center}
\end{figure}

Figure~\ref{fig:ene_diff_qi_ptmc} compares the ground-state energy 
obtained by the QI search method and the PTMC method 
for 120 different instances of the random coupling realizations in $\hat{H}_0$ with the system sizes $N=200$, $600$, and $1000$. 
For the PTMC calculations, we adopt a large number of MC sweeps ($2 \times 10^5$) 
and a comparable numerical condition described in Ref.~\cite{Rozada2019}.
As shown in the figure, for the system size $N=200$, 
the QI search with the input parameters $(L, K, I)=(N,N,N)$ finds the same ground-state energies as the PTMC method 
for 95 random coupling instances out of 120 instances, while for the remaining 25 random coupling instances the QI search is 
trapped in metastable states with larger energies. 

For $N=600$, the success rate of the QI search finding the same ground-state energy as the PTMC method 
is down to $6/120$ with the input parameters $(L, K, I)=(N,N,N)$, 
$9/120$ with $(N,2N,2N)$, $18/120$ with $(N,5N,5N)$, and $26/120$ with $(N,10N,10N)$, 
showing a gradual improvement with increasing $K$ and $I$. 
On the other hand, in the case of 
$N=1000$, the QI search can not find the ground-state energy consistent with the PTMC method 
for all the 120 random coupling realizations, even when we set the input parameters $(L, K, I)=(N,10N,10N)$. 
These results suggest that further increase of $(L,K,I)$ or better tuning of the form of $\hat{H}_1$ is required to 
efficiently search the ground state in large systems.

\clearpage

\input{random_20210716_v2_v2_HU.bbl}

\end{document}

%% file: random_20210716_v2_v2_HU.bbl
%